# Information Loss as a Foundational Principle for the Second Law of Thermodynamics

**T. L. Duncan**
Center for Science Education, Portland State University, Portland, Oregon 97207-0751, USA and Department of Physics, Pacific University, 2043 College Way, Forest Grove, Oregon 97116, USA. E-mail: duncant@pdx.edu.

**J. S. Semura**
Department of Physics, Portland State University, Portland, Oregon 97207-0751, USA. E-mail: semuraj@pdx.edu.

**Key words:** second law, entropy, information, statistical mechanics, thermodynamics

## Abstract

In a previous paper [1] we considered the question, "What underlying property of nature is responsible for the second law?" A simple answer can be stated in terms of information: The fundamental loss of information gives rise to the second law. This line of thinking highlights the existence of two independent but coupled sets of laws: Information dynamics and energy dynamics. The distinction helps shed light on certain foundational questions in statistical mechanics. For example, the confusion surrounding previous "derivations" of the second law from energy dynamics can be resolved by noting that such derivations incorporate one or more assumptions that correspond to the loss of information. In this paper we further develop and explore the perspective in which the second law is fundamentally a law of information dynamics.

## 1. Introduction

The confusion and controversy which have surrounded the second law of thermodynamics throughout its history demonstrate the subtlety underlying the fundamental essence of the law. Despite widespread confidence among physicists in the universal validity of the second law, it is striking that there is no corresponding agreement on the justification for why the second law holds, nor even exactly what it says. Such clarity is desirable from the standpoint of understanding nature, but is even more important in light of the growing string of challenges that stretch the limits of what it means for the law to hold or be violated [2, 3]. In order to address these issues, it is important to have a clear formulation of exactly what fundamental restriction the second law places on natural processes. An indication that this clarity has been achieved would be a straightforward answer to the question, "What property of nature would have to be changed in order to 'turn off' the second law?"

In the rest of this paper, we further develop the idea introduced in [1], that a direction for achieving this clarity is to view the second law as fundamentally a law of information dynamics. In Section 2, we review some of the background and motivation for this connection between entropy and information. In Section 3, we argue that the identification of information dynamics as separate from (and additional to) energy dynamics resolves puzzles related to "deriving" the irreversible second law from time-symmetric energy dynamics. In Section 4, we present our perspective in summary form. In Section 5 we illustrate the usefulness of this perspective on the second law and discuss how the perspective may fit within a broader context of information dynamics in physics.

## 2. Background on the Entropy-Information Connection

The general idea of connecting thermodynamic entropy to information is widely known, but it is not a connection everyone agrees is valid. The original thermodynamic entropy defined by Clausius in the first rigorous formulation of the second law [4] has no obvious connection to information at all. His thermodynamic entropy, $S_C$, is defined by the differential expression $dS_C = \frac{dQ_{rev}}{T}$, where $Q_{rev}$ is the heat transferred to the system in a reversible process, and $T$ is the temperature (in Kelvin) at which the heat transfer occurs. With this definition, Clausius formulated the second law concisely with the statement, "The entropy ($S_C$) of the universe tends toward a maximum."

Boltzmann's microscopic interpretation of entropy [5] opened the way for a connection to information. The possibility of a connection is made more apparent by the formal mathematical equivalence of Shannon information entropy and (for example) the Gibbs expression for statistical mechanical entropy:

$$S_{SI} = -K \sum_i p_i \log_2 p_i \qquad (1)$$

$$S_G = -k_B \sum_i p_i \ln p_i, \qquad (2)$$

where K is an arbitrary constant and $k_B$ is Boltzmann's constant. But, as ter Haar [6] and others have pointed out, this mathematical connection could be misleading and does not necessarily require a deeper link.

A strong argument that the information-entropy connection *is* fundamental came from Jaynes [7,8]. He showed that familiar thermodynamic results emerge naturally from an information theoretic treatment. A key aspect of Jaynes' approach is that it is not in any way an attempt to derive statistical mechanics from the underlying microscopic physics. Rather, it is a subjective method of inference based on the observer's limited knowledge of the state of the system. Jaynes showed that the familiar laws of statistical mechanics emerge from treating our ignorance in an unbiased way. Thus this approach still leaves open the question of what microscopic dynamics is "really" going on (and how it conspires to match the information theory result).

Another line of thought connecting entropy and information is related to Szilard's 1929 thought experiment involving Maxwell's Demon [9]. Szilard's example highlights the physical nature of information, which is important to our argument. His thought experiment led to the realization that the seemingly abstract concept of information *must* be included in the accounting of entropy if the second law is to be universal. Otherwise, if there is no entropy generation associated in some way with handling information about the state of the system, then a Maxwell Demon could defeat the second law by sorting molecules based on the information gathered. This perspective was later clarified in more detail by others, including Landauer [10], Bennett [11,12], and Penrose [13], who pointed out that erasing (more accurately, *resetting* to a standard state – see *e.g.* the discussion in [14]) is the key step in information handling that links physical entropy and information. This has become known as Landauer's Principle: Resetting ("erasing") 1 bit of information increases the entropy of the environment by at least $k_B \ln 2$ (*i.e.* resetting is an inherently dissipative process). The bottom line from this perspective is that one is faced with a choice: either the kind of abstract information we deal with intuitively is subject to the laws of thermodynamics; or if information is independent of thermodynamic restrictions, then it can be used to violate the second law. So, since the familiar kind of information is subject to second law, maybe the second law is fundamentally all about information.

## 3. Does Information Play a Necessary Role?

At this point one could still argue that the information picture is convenient and useful for understanding, but ultimately superfluous – everything it explains could also be fully justified without mention of information at all. Consider by analogy the center of mass description of a collection of individual particles. Often it is convenient to describe only the motion of the center of mass, rather than the detailed motions of the individual particles. But in this case it is clear that the center of mass description is derivative, not fundamental. The center of mass description is convenient, but there is nothing fundamentally new in it that was not already contained in the individual particle description.

The crux of the idea we're exploring is that there is an information dynamics which is separate from energy dynamics, and is *not* derivative in the way the center of mass description is. It amounts to taking very seriously the clue from Jaynes' analysis: Within the information framework, statistical mechanics follows naturally with no additional arbitrary assumptions. By contrast, one of the long-standing problems in the foundations of statistical mechanics without information dynamics is to explain why statistical mechanics works so well. Why do the supposedly more fundamental microscopic dynamics produce the same behavior as the information treatment described by Jaynes and others?

Perhaps the strongest argument for the reality of information dynamics comes from attempts to derive the second law without reference to information, based only on the microscopic energy dynamics of classical Hamiltonian/Lagrangian mechanics or unitary quantum mechanics. Different derivations involve different assumptions and approximations. But the common theme to "successful derivations" of the second law is that they involve at least one approximation that effectively *loses information* [15]. For example, a stochastic or random phase approximation throws away information about the exact state of a system before a collision. Coarse graining loses information by reducing the number of variables. Quantum mechanical decoherence also can play the role of a mechanism which effectively loses information.

The key insight here is that *when one attempts to derive the second law without any reference to information, a step which can be described as information loss always makes its way into the derivation by some sleight of hand.* An information-losing approximation is necessary, and adds essentially new physics into the model which is outside the realm of energy dynamics. It must be outside, because Liouville's theorem (or its quantum version) demonstrates the phase space volume-preserving nature of time evolution under energy dynamics, which prevents this dynamics from losing information. The point is that you must lose information in order to make the leap from microscopic energy dynamics to the second law. This seems to be an important insight: Since our models throw away information in order to match with what we observe in nature, this suggests that information really is lost in nature.

## 4. Summary of the Perspective

Given this background, it seems natural to consider the possibility that there is a fundamental, objective dynamics of information that we should take seriously as underlying the second law. The success of Jaynes' analysis makes sense if the reason it works so well is that it taps into an objective dynamics of information that is just as fundamental as the familiar laws of energy dynamics. Rather than just saying nature behaves as if information is lost, what if it really is lost in a deep and fundamental sense? The perspective we suggest is worth exploring can be summarized with the following points:

1) Energy and information dynamics are independent but coupled (see Figure 1).

2) The second law of thermodynamic is not reducible purely to mechanics (classical or quantum); it is part of information dynamics. That is, the second law exists because there is a restriction applying to information that is outside of and additional to the laws of classical or quantum mechanics.

3) The foundational principle underlying the second law can then be expressed succinctly in terms of information loss:

> "*No process can result in a net gain of information.*"

In other words, the uncertainty about the detailed state of a system cannot decrease over time – uncertainty increases or stays the same. It is important to note that this need not be a subjective statement. The uncertainty can be real. If the state of a system is truly not any more well-defined than the constraints that have been placed on it, then Jaynes' analysis applies to the real state of the system, not just a particular observer's knowledge of that system.

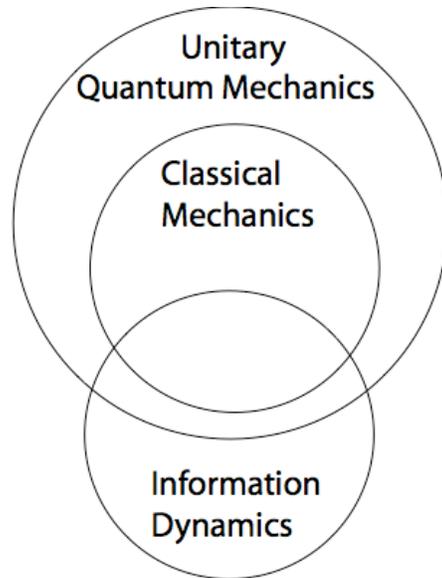

**Figure 1** Information dynamics provides additional laws governing the behavior of a system which are not contained within Hamiltonian/Lagrangian classical or quantum mechanics.

From this perspective, the existence of the second law of thermodynamics and the sleight of hand necessary to derive it from energy dynamics are expressions of the fact that its foundation lies within information dynamics. In essence, it could explain why Jaynes' approach of maximizing information entropy subject to the known constraints on the system works so well: because that is what nature is actually doing at the fundamental level.

## 5. Discussion and Speculation

What is the benefit of this perspective on the second law? It should be noted that expressing the law in terms of information loss does *not* explain the second law. But it does focus our attention on understanding the dynamics of information (tracking its flow and where it is lost or gained) in order to understand the basis of the second law. The question, "Why is there a second law of thermodynamics?" now becomes, "Why is information lost with time?"

This phrasing of the question connects the subject to work exploring the foundations of quantum mechanics in terms of information (see *e.g.* [16-20]), and raises the interesting possibility that fundamental aspects of quantum mechanics and statistical mechanics might both arise from different aspects of underlying information laws.

The information loss perspective provides a natural framework for incorporating extensions and apparent challenges to the second law. The principle that "no process can result in a net gain of information" appears to be deeper and more universal than standard formulations of the second law. For example, consider a classical heat engine, which requires both a high temperature heat source ($T_H$) and a low temperature heat sink ($T_C$) in order to operate. From the information perspective, the $T_C$ sink is required in order to provide a *loss* of information to compensate for the *gain* in information associated with heat flowing out of the $T_H$ source and doing macroscopic work (W). That is, the state of the $T_H$ and W component of the system becomes more well-defined, but the state of the $T_C$ sink becomes less well-defined, so that the total system becomes less well-defined (or its information content remains constant, in the thermodynamically reversible case).

This analysis within the information framework is equivalent to standard thermodynamics for the classical heat engine. But it is more general, since it doesn't specifically require a low temperature heat sink. Rather, it requires that *information* must be lost somewhere else if it is gained in one place. This allows the framework to incorporate new possibilities for storage and loss of information that go beyond standard thermodynamics. For example, quantum coherence and entanglement (which are involved in many of the recent second law challenges) provide situations where information can be stored in unrecognized channels. (These are Class II challenges in the categorization proposed in [1].)

Consider the quantum heat engine (QHE) described by Scully [21]. Since such a heat engine does not require a low temperature sink in order to operate, it might appear superficially inconsistent with some formulations of the second law (*e.g.* it can exceed the classical Carnot efficiency). However, from the information perspective, the more general requirement is only that information must be lost somewhere else if it is gained in one place. There can be other ways to lose information besides heat flow into a low temperature sink. For the QHE, information loss occurs through the spreading out of the wave packet associated with the atom which serves as the working fluid for the heat engine. In other words, the QHE can operate without a cold sink because it has another outlet for losing information; namely, the location of the atom, which becomes less well-defined in each cycle of the engine.

Similar analyses may provide insight into other examples which stretch the meaning of the second law (*e.g.* [22, 23]). Class II challenges involving quantum entanglement [23] are particularly interesting because in these cases information is non-local (shared jointly among distinct parts of the system). This offers another mechanism for apparent second law violations, because entangled information is not accounted for in the usual thermodynamic bookkeeping where parts of the system are treated separately. A system which contains entangled information thus has another channel for information loss in addition to the usual channel of heat dissipation. Allahverdyan and Nieuwenhuizen [23] provide a nice example of this effect, where entanglement allows violation of the classical Clausius inequality (without violating the Thomson "no net work in a complete cycle" formulation of the second law), and leads to a breakdown of the usual Landauer bound. Again, in the context of our information-loss framework, this can be understood as a case where the usually required dissipation of heat has been replaced by loss of information that was stored by entanglement.

These examples may be just the beginning, suggesting that the information-loss framework offers the possibility of discovering new mechanisms of information storage through the analysis of second law challenges, deepening our understanding both of the second law and of information dynamics.

**Acknowledgements:** The authors would like to dedicate this work to Patty Jeanne Semura. We also thank D. P. Sheehan for organizing this AAAS symposium exploring the foundations and status of the second law.